\documentclass[twocolumn,showpacs,prl]{revtex4-1}

\usepackage{graphicx}
\usepackage{dcolumn}
\usepackage{bm}

\begin{document}

\title{Haldane state with toroidal magnetic order}

\author{V.I.~Belyavsky and Yu.V.~Kopaev}

\affiliation{P.N.~Lebedev Physical Institute of Russian Academy of
Sciences, \\  Moscow, 119991, Russia}


\begin{abstract}

We show that a simplified two-band model describing toroidal
magnetic order in two-dimensional crystal is entirely equivalent
to the well-known Haldane model of a honeycomb lattice in periodic
internal magnetic field with zero total flux through the unit
cell. Crystals with toroidal ordering can be considered as real
physical systems which should exhibit topologically nontrivial
state, predicted by Haldane, that is similar to the quantum Hall
effect regime without any external magnetic field. Analogous
description is formulated for quantum spin Hall effect without any
spin-orbit interaction.

\end{abstract}


\maketitle

Highly speculative model of a two-dimensional (2D) honeycomb
lattice with two nonequivalent sublattices in periodic internal
magnetic field, introduced by Haldane more than twenty years ago
\cite{Haldane}, directly predicts a new state of matter similar to
the integer-valued Quantum Hall Effect (QHE) regime without any
external magnetic field. It is presupposed in this Haldane model
that internal magnetic-flux density with full symmetry of 2D
lattice results in zero total flux through the unit cell. In the
framework of the Haldane model, time-reversal invariance and 2D
space-inversion symmetry are broken by the internal magnetic field
and the non-equivalence of the sublattices, respectively.
Tight-binding model \cite{Haldane} takes into account hopping
matrix elements between nearest and second-neighbor sites (the
nearest sites of a sublattice) and the periodic vector potential
${\bm{A}}({\bm{r}})$ of the internal field is included into this
model through the phase multipliers $\exp{[i(e/{\hbar}c) \int
{\bm{A}}d{\bm{r}}]}$ of the hopping matrix elements. Thus,
magnetic field in the model \cite{Haldane} becomes apparent
through the vector potential in the phase factors of the wave
functions similar to that in Bohm-Aharonov effect. Since internal
magnetic field breaks time-reversal invariance, quasiparticle
spectrum of the system with broken spatial inversion turns out to
be nonsymmetric with respect to inversion of quasi-momentum,
${\bm{k}}\leftrightarrow -{\bm{k}}$. In addition, edge chiral
fermion excitations, typical of the QHE regime, arise without
partners of opposite chirality.

It should be noted that, simultaneously with and independently of
the paper Ref.\cite{Haldane}, Volovik \cite{V1,V2} has predicted
and studied in detail the half-integer QHE in the A-phase of
liquid helium-3 without both external magnetic field and rotation
of the container.

In the QHE regime, quantizing external magnetic field normal to
the plane of 2D crystal both transforms 2D metal into 2D insulator
in which degenerated Landau levels play role of the bands of
allowed states and also determine a direction of a drift of the
cyclotron orbit center along the boundary of the crystal. Finite
distance between the neighboring Landau levels and the absence of
the backscattering in the edge chiral states result in a stability
of the ground state of the QHE regime with respect to relatively
weak structure perturbations. Comparatively weak internal periodic
magnetic field in the Haldane model breaks time-reversal symmetry
but in itself cannot create an insulating state in the form of a
system of Landau levels like that in the QHE regime. However, as
was also proposed by Haldane \cite{Haldane}, sometimes the
internal field could be associated with a magnetic ordered state
with broken time-reversal invariance that might result in an
insulating gap in the quasiparticle spectrum. Finite insulating
gap and absence of backscattering could lead to a robust Haldane
state similar to the QHE regime.

Thouless et al. \cite{TKNN} have shown that QHE regime differs
from the conventional band insulating state by the integer-valued
topological invariant: in the case of conventional (trivial)
insulator, such a topological invariant turns out to be zero,
$n=0$ and edge states are absent; $n\neq 0$ may be referred to a {
\it{topological insulator}}.

Edge states can also arise in the topological insulator with
spin-orbit interaction which does not break time-reversal symmetry
\cite{KM,BHZ}. Such edge states correspond to a transfer not of
electric charge but of spin. Electrons with opposite spin
projections move in opposite directions along crystal boundaries.
This becomes apparent as spin quantum Hall effect \cite{KM,BHZ}.

As it was pointed out by Haldane \cite{Haldane} and Volovik
\cite{V2}, topologically nontrivial insulating state with electric
charge transfer over the edge states could be realized in 2D
systems with magnetic order breaking time-reversal invariance of
the Hamiltonian. However, one can conclude that such states have
never been experimentally realized till now. In this Letter, we
show that two-band model of excitonic insulator \cite{KK} with
electron-hole pairing leading to toroidal magnetic order turns out
to be equivalent to the Haldane model after very natural
simplification. We believe the Haldane state can be realized just
in 2D insulating crystals with toroidal magnetic order.

Such insulating crystals with spontaneous toroidal moment of the
unit cell in the ordered phase are investigated theoretically long
ago \cite{VGKT}. Since then, a great number of crystals and
artificial 2D heterostructures with toroidal orderring were
discovered \cite{Spaldin} mainly because of their relationship to
the magnetoelectric effect.

Toroidal magnetic moment ${\bm{t}}$ arises in the multipole
expansion of the electrodynamic vector potential
${\bm{A}}({\bm{r}})$ beginning with the second order term which
can be presented as a sum of the contributions from the magnetic
quadrupole and toroidal moments \cite{Dubovik}. The toroidal part
of the vector potential created by orbital currents at some
distant point ${\bm{r}}$ can be written as \cite{Spaldin}
\begin{equation}\label{1}
{\bm{A}}({\bm{r}})=\nabla
({\bm{t}}\nabla)r_{}^{-1}+4{\pi}{\bm{\tau}}({\bm{r}}),
\end{equation}
where
\begin{equation}\label{2}
{\bm{\tau}}({\bm{r}})={\frac{1}{6c}}[{\bm{r}}\times [{\bm{r}}
\times {\bm{j}}({\bm{r}})]]
\end{equation}
is the toroidal moment density, ${\bm{j}}({\bm{r}})$ is the
orbital electric current density. Nonzero toroidal moment
corresponds to a special distribution of orbital currents in
Eq.~({\ref{2}). As a classical example, one can consider the
so-called {\it{poloidal}} currents flowing through a solenoid that
is bent into a torus.

One can imagine that all orbital currents are shrunk into the
origin of coordinates \cite{Spaldin}. Then the toroidal moment
density (\ref{1}) should be written as ${\bm{\tau}}\rightarrow
{\bm{t}}\cdot{\delta}({\bm{r}})$. The first term in Eq.~(\ref{1})
can be eliminated by appropriate gauge transformation so that the
vector potential takes the form ${\bm{A}}({\bm{r}})=
4{\pi}{\bm{\tau}}({\bm{r}})$.

It should be noted that toroidal moment can be created not only by
persistent orbital currents but also by an appropriate
configuration of magnetic dipole moments, such as head-to-tail
arrangement of spins \cite{Spaldin}.

One can see that the toroidal contributions into both vector
potential and magnetic flux density
${\bm{B}}={\text{rot}}{\bm{A}}$ turn out to be localized inside a
bounded domain of the real space in which toroidal moment density
is nonzero. In the case of crystalline solid, such a domain is the
unit cell so that both vector potential and magnetic flux density
should be periodic functions with basic periods of the solid
similar to that proposed in the Haldane model.

Since toroidal moment density ${\bm{\tau}}({\bm{r}})$ is $t$-odd
polar (${\bm{r}}$-odd) vector, the mean field order parameter
originating from toroidal ordering should change sign upon both
time reversal and spatial inversion. As direct product of electric
and magnetic field strengthes has the same symmetry toroidal order
may be apparent as a linear magneto-electric effect. Corresponding
contribution into the free energy of the form
$F^{}_{me}=-{\alpha}^{}_{ik}E^{}_iH^{}_k$ where ${\alpha}^{}_{ik}$
is a magneto-electric tensor. Magnetic field induced electric
polarization and electric field induced magnetization are
determined by this tensor as $P^{}_i= {\alpha}^{}_{ik}H^{}_k$ and
$M^{}_i= {\alpha}^{}_{ki}E^{}_k$, respectively.

From the symmetry consideration, magneto-electricity should be one
of the distinctive properties of the toroidal ordered state
\cite{Spaldin}: ${\bm{P}}=-[{\bm{t}}\times {\bm{H}}]$,
${\bm{M}}=[{\bm{t}}\times {\bm{E}}]$. Magneto-electric tensor can
be decomposed into a pseudoscalar, a polar vector and a symmetric
traceless tensor. Thus, non-diagonal components of the
magneto-electric tensor should be nonzero and antisymmetric,
${\alpha}^{}_{ik}=-{\alpha}^{}_{ki}$ \cite{GKT}.

The bulk adiabatic magneto-electric tensor determines the surface
Hall impedance in a magneto-electric crystal \cite{Widom}, $
R^{-1}_{H}=c\, n^{}_i {\alpha}^{}_{ik}n^{}_k$, where $c$ is light
velocity. This thermodynamic relation is used in Ref.
\cite{Haldane} to calculate quantized Hall conductivity
${\sigma}^{}_{xy}$.

A rise of the toroidal ordering can be a result of an electron
phase transition. It is convenient to describe such a transition
in the framework of the model of the excitonic insulator with
high-temperature (more symmetric) phase being 2D semimetal with
equally centered isotropic electron and hole bands \cite{VGKT}.

Bloch Hamiltonian of such two-band semimetal can be written as
\begin{equation}\label{3}
{\hat{H}}^{}_{\bm{p}}={\varepsilon}^{}_+({\bm{p}})\,{\hat{\tau}}_{}^{0}+
{\varepsilon}^{}_-({\bm{p}})\,{\hat{\tau}}_{}^{3}
+({\bm{\gamma}}{\bm{p}})\,{\hat{\tau}}_{}^{2},
\end{equation}
where ${\hat{\tau}}_{}^{0}$ is the unit matrix,
${\hat{\tau}_{}^{1,2,3}}$ are the Pauli matrixes,
$2{\varepsilon}^{}_{\pm}({\bm{p}})={\varepsilon}^{}_1({\bm{p}})\pm
{\varepsilon}^{}_2({\bm{p}})$, ${\varepsilon}^{}_s({\bm{p}})$ is
electron dispersion in the band $s$ (electron energy is counted
from the chemical potential), ${\bm{\gamma}}={\bm{p}}^{}_{12}/m$
is a real vector, ${\bm{p}}^{}_{12}$ is interband matrix element
of momentum, $m$ is electron mass. Note that ${\bm{p}}^{}_{12}\neq
0$ when, in particular, the bands of the semimetal are formed by
the hybrid states of different parity, for example, $s$ and $p$
states.

Fermi contour (FC), that is a boundary separating filling and
vacant states in the electron and hole bands, is determined by the
Fermi momentum ${\bm{p}}^{}_F$. In the case of isotropic bands the
FC is a circle with radius $p^{}_F$. One can choose $p^{}_1$-axis
along the vector ${\bm{\gamma}}$. For low-energy elementary
excitations a deviation ${\bm{k}}$ of quasimomentum from the Fermi
momentum ${\bm{p}}^{}_F$ is small. Therefore, in the first
approximation, one can replace the quasimomentum ${\bm{p}}$ with
${\bm{p}}^{}_F$ in the non-diagonal terms of Hamiltonian (\ref{3})
whereas the diagonal terms depend on ${\bm{k}}$ directly because
energies ${\varepsilon}^{}_1({\bm{p}})$ and
${\varepsilon}^{}_2({\bm{p}})$ are counted from the chemical
potential.

Hamiltonian (\ref{3}) is characterized by gapless anisotropic
quasiparticle spectrum
\begin{equation}\label{4}
E^{}_{1,2}({\bm{k}})=
{\varepsilon}^{}_{+}({\bm{k}})\pm{\sqrt{{\varepsilon}^2_{-}
({\bm{k}})+({\bm{\gamma}}{\bm{p}}^{}_F)_{}^2}},
\end{equation}
which is symmetrical with respect to quasimomentum inversion,
${\bm{k}}\rightarrow -{\bm{k}}$.

To obtain both time-reversal and space-inversion symmetries to be
broken one can consider the low-temperature ordered phase
originating from electron-hole pairing resulting in nonzero
toroidal moment. The order parameter is determined by an anomalous
average $\langle {\hat{c}}^{\dag}_{1{\bm{p}}{\sigma}}
{\hat{c}}^{}_{2{\bm{p}}{\sigma}_{}^{\prime}}\rangle$ corresponding
to singlet (${\sigma}={\sigma}_{}^{\prime}$) or triplet pairing of
electron and hole belonging to different bands. Here,
${\hat{c}}^{}_{s{\bm{p}}{\sigma}}$ annihilates electron with
quasimomentum ${\bm{p}}$ and spin $\sigma$ in the band $s=1,2$.

It is well known that orbital and spin current states
corresponding to singlet and triplet electron-hole pairing,
respectively, can result in the ordered states with pure imaginal
order parameter \cite{HR}.

We restrict ourselves to the case of singlet electron-hole pairing
resulting in a pure imaginal order parameter $i{\Delta}(\bm{k})$
where ${\Delta}(\bm{k})$ is a real function of quasimomentum
${\bm{k}}$. In such a case, toroidal order arises from poloidal
charge currents. To associate the order parameter with the
toroidal moment one should use the relation ${\bm{t}}=
{\bm{\gamma}}{\Delta}$ following from the interrelation between
the charge current density and toroidal moment of the unit cell
\cite{Kopaev}.

The singularity in the electron-hole pairing channel that could
lead to the insulating ordered state at low temperature can arise
under nesting of electron dispersion,
${\varepsilon}^{}_1({\bm{k}})=-{\varepsilon}^{}_2({\bm{k}})\equiv
{\varepsilon}({\bm{k}})$. We suppose that such a condition is
fulfilled in a vicinity of the FC. Then the mean field Bloch
Hamiltonian corresponding to singlet electron-hole pairing with a
rise of a toroidal magnetic moment of the unit cell and pure
imaginal order parameter ${\Delta}({\bm{k}})$ takes the form
\begin{equation}\label{5}
{\hat{H}}^{}_{\bm{k}}={\hbar}({\bm{v}}^{}_F{\bm{k}})\,{\hat{\tau}}_{}^{3}
+[({\bm{\gamma}}{\bm{p}}^{}_F)+{\Delta}({\bm{k}})]\,{\hat{\tau}}_{}^{2},
\end{equation}
where electron dispersion is presented by linear function of
quasimomentum, ${\varepsilon}({\bm{k}})=
{\hbar}({\bm{v}}^{}_F{\bm{k}})$. Here, ${\bm{v}}^{}_F$ is Fermi
velocity. Later on, we restrict ourselves to the case of isotropic
($s$-wave) toroidal order parameter ${\Delta}={\text{const}}$.

The spectrum of the high-temperature phase, when ${\Delta}=0$,
\begin{equation}\label{6}
E^{}_{1,2}({\bm{k}})=\pm
{\sqrt{{\hbar}_{}^2({\bm{v}}^{}_F{\bm{k}})_{}^2+
({\bm{\gamma}}{\bm{p}}^{}_F)_{}^2}},
\end{equation}
turns out to be gapless because of the fact that ${\bm{\gamma}}
{\bm{p}}^{}_F=0$ for two antipodal directions of the
$p^{}_2$-axis. Two points on the FC corresponding to such
directions form two Dirac cones with linear dispersion at small
${\bm{k}}$.

In the low-temperature phase, when ${\Delta}\neq 0$, quasiparticle
spectrum corresponding to Bloch Hamiltonian (\ref{5}) becomes
gapped,
\begin{equation}\label{7}
E^{}_{1,2}({\bm{k}})=
\pm{\sqrt{{\hbar}_{}^2({\bm{v}}^{}_F{\bm{k}})_{}^2+
({\bm{\gamma}}{\bm{p}}^{}_F+{\Delta})_{}^2}}.
\end{equation}
Similar to the Haldane model \cite{Haldane}, the spectrum turns
out to be asymmetric with respect to ${\bm{k}}\rightarrow
-{\bm{k}}$: $E^{}_{1,2}(-{\bm{k}})\neq E^{}_{1,2}({\bm{k}})$.

Note that quasiparticle dispersion (\ref{7}) depends on ${\bm{k}}$
not only over the first term under the square root but also over
the scalar product depending on quasimomentum polar angle
${\varphi}$. Since vector ${\bm{\gamma}}$ is directed along
$p^{}_1$ axis, $({\bm{\gamma}}{\bm{p}}^{}_F)>0$ at
$|{\varphi}|<{\pi}/2$; in the opposite case,
$({\bm{\gamma}}{\bm{p}}^{}_F)<0$.

The energy gap in the spectrum (\ref{7}) depends on quasimomentum
considerably. In particular, it equals to $2{\Delta}$ at the
points of intersection of the FC and $p^{}_2$-axis. In the case of
${\bm{\gamma}}{\bm{p}}^{}_F> 0$, the gap monotonically increases
from $2{\Delta}$ when polar angle ${\varphi}$ between
${\bm{\gamma}}$ and ${\bm{p}}^{}_F$ varies from $\pm {\pi}/2$
(these points corresponds to ${\bm{\gamma}}{\bm{p}}^{}_F=0$) to
zero (if ${\varphi}=0$, the gap equals to
$2{\sqrt{{\gamma}_{}^2p^2_F+{\Delta}}}$).

If $({\bm{\gamma}}{\bm{p}}^{}_F)<0$, the energy gap is a
nonmonotone function of the polar angle. It vanishes at the points
of the FC with polar angle ${\varphi}= \pm {\varphi}_{c}^{}$
corresponding to ${\bm{\gamma}}{\bm{p}}^{}_F+{\Delta}=0$. Thus, if
${\Delta}$ is small enough the spectrum (\ref{7}) remains gapless.
The gap can arise in those directions of the momentum space where
the toroidal order parameter exceeds ${\gamma}p^{}_F$:
${\Delta}>{\gamma}p^{}_F$.

In the case ${\Delta}<{\gamma}p^{}_F$, one can expect that Bloch
Hamiltonian (\ref{5}) should be related to a semimetal with
insulating gap on some part of the FC corresponding to
$-1<{\cos{\varphi}}<-{\Delta}/{\gamma}p^{}_F$. Distribution of
semimetal and insulating constituents in the real space
corresponding to a peculiar intermediate state may be highly
complicated. One can think that the phase transition between
semimetal and insulating phases should be occurred under the
condition that the insulating toroidal order parameter exceeds
several critical value, ${\Delta}>{\Delta}_{c}^{}$, where
${\Delta}_{c}^{}$ can be considered as a minimum value of the
toroidal order parameter.

Thus, owing to ${\varphi}$-dependence, the mean field Bloch
Hamiltonian (\ref{5}) describes rather complicated low-temperature
insulating phase with toroidal magnetic order. Hamiltonian
(\ref{5}) can be simplified considerably if one replaces scalar
product ${\bm{\gamma}}{\bm{p}}^{}_F$ with constants corresponding
to the averages over the polar angle in the limits of the domains
of constant sign of this scalar product,
${\bm{\gamma}}{\bm{p}}^{}_F\rightarrow
{\alpha}{\gamma}_{c}^{}p^{}_F$. Here
${\alpha}={\text{sgn}}({\bm{\gamma}}{\bm{p}}^{}_F)$,
${\gamma}_{c}^{}={\Delta}_{c}^{}/p^{}_F$.

In just the same way as in the case of the Haldane model, mean
field Bloch Hamiltonian (\ref{5}) can be reduced to two
independent effective Hamiltonians,
\begin{equation}\label{8}
{\hat{H}}^{}_{{\alpha}{\bm{k}}}={\hbar}({\bm{v}}^{}_F{\bm{k}}){\hat{\tau}}_{}^{3}
+[{\alpha}{\gamma}_{c}^{}p^{}_F+{\Delta}]{\hat{\tau}}_{}^{2},
\end{equation}
which are distinguished by discrete parameter ${\alpha}=\pm 1$,
which connects the terms breaking space inversion and time
reversal symmetries
$({\bm{\gamma}}{\bm{p}}^{}_F)+{\Delta}\rightarrow
{\alpha}{\gamma}_{c}^{}p^{}_F+{\Delta}$.

One can see a direct analogy between the pairs of effective Bloch
Hamiltonians corresponding to extremely simplified problem of the
excitonic insulator with toroidal magnetic order (\ref{8}) and the
Haldane model \cite{Haldane} described by the effective
Hamiltonians (we use somewhat other designations with respect to
the original paper \cite{Haldane})
\begin{equation}\label{9}
{\hat{H}}_{{\alpha}{\bm{k}}}^{(0)}={\hbar}v^{}_F(k^{}_x{\tau}_{}^2-k^{}_y{\tau}_{}^1)+
m^{}_{\alpha}v^2_F{\tau}_{}^3
\end{equation}
which can be considered as a presentation of a traceless $2\times
2$ matrices in the form of a linear combination of the Pauli
matrices. The two effective Hamiltonians (\ref{8}) coincide with
(\ref{9}) nearly verbally if one considers more general case of
the ordering with nonzero both real and imaginal parts of the
order parameter,
${\Delta}={\Delta}_{}^{\prime}+i{\Delta}_{}^{\prime
\prime}=|{\Delta}|\, e_{}^{i{\phi}}$. Here, ${\phi}$ is a phase of
the order parameter (${\phi}={\pi}/2$ in the case of pure imaginal
${\Delta}$).

The antipodal directions of quasimomentum, differing with sign of
$\alpha$, turn out to be considerably nonequivalent at
${\Delta}\neq 0$. Energy gap
$E^{}_{g+}=2({\Delta}_{c}^{}+{\Delta})$ is an increasing function
of  $\Delta$ in the case of a direction corresponding to
${\alpha}=+1$, whereas the gap
$E^{}_{g-}=2|{\Delta}_{c}^{}-{\Delta}|$ for a direction
corresponding to ${\alpha}=-1$ exhibits a minimum at
${\Delta}={\Delta}_{c}^{}$ in which $E^{}_{g-}=0$. Change of sign
of the gap corresponds to the band inversion similar to that in
the systems with quantum spin Hall effect \cite{BHZ}. It is
important that band inversion in excitonic insulator is not a
result of the spin-orbit interaction so that one can expect that
the Haldane state can be observed at rather high temperature.

Approximate Hamiltonians  ${\hat{H}}^{}_{{\alpha}{\bm{k}}}$
corresponding to two-band model of excitonic insulator with
toroidal magnetic order and effective Hamiltonians
${\hat{H}}^{(0)}_{{{\alpha}}{\bm{k}}}$ of the Haldane model
describing 2D honeycomb lattice are not identical. Nevertheless,
both ${\hat{H}}^{(0)}_{{{\alpha}}{\bm{k}}}$ and
${\hat{H}}^{}_{{{\alpha}}{\bm{k}}}$ result in the same structure
of quasiparticle spectrum. In the case of excitonic insulator we
have
\begin{equation}\label{10}
E^{\alpha}_{1,2}({\bm{k}})=
\pm{\sqrt{{\hbar}_{}^2({\bm{v}}^{}_F{\bm{k}})_{}^2+
({\alpha}{\Delta}_{c}^{}+{\Delta})_{}^2}},
\end{equation}
where the second term under the square root plays role of the mass
term $m^{}_{\alpha}$ defined in the Haldane model \cite{Haldane}
as
$$m^{}_{\alpha}\equiv M-3{\sqrt{3}}{\alpha}t^{}_2 {\sin{\phi}}
\rightarrow {\alpha}{\Delta}^{}_c+|{\Delta}|{\sin{\phi}};$$ here
$\pm M$ are the energies of the sites of two different sublattices
of the honeycomb lattice, $t^{}_2$ is a matrix element
corresponding to the hopping between the second neighbor sites,
$\phi$ is an additional phase corresponding to such a hopping in
the Haldane model \cite{Haldane}.

Note that triplet electron-hole pairing described by pure imaginal
order parameter can be associated with poloidal spin currents in
the volume \cite{GK} and therefore should be apparent in spin
quantum Hall effect. Triplet pairing can be described
approximately by the Hamiltonian in the form of a direct sum of
two pairs of effective Hamiltonians
${\hat{H}}^{}_{{\bm{k}}{\sigma}{\alpha}}$, similar to (\ref{8}) or
(\ref{9}), for different spin projections $\sigma$, where
${\alpha}=\pm 1$ for $\sigma =\uparrow$ whereas ${\alpha}=\mp 1$
for $\sigma =\downarrow$. Similar model was introduced by
Bernevig, Huges and Zhang \cite{BHZ} to study the quantum spin
Hall effect in HgTe/CdTe quantum wells.

Any Hermitian operator ${\hat{H}}_{}^{(0)}$ is fully determined by
its spectrum $E^{}_{n}$ and a complete set of eigenfunctions
${\psi}^{}_{n{\nu}}$, where quantum number $\nu$ takes into
account a degeneracy of the eigenvalue $E^{}_n$. Such a set can be
used as an orthonormal basis of the Hilbert space in which
operator ${\hat{H}}_{}^{(0)}$ acts. Let us consider another
Hermitian operator ${\hat{H}}_{}^{}$ acting in the same space and
suppose that ${\hat{H}}_{}^{(0)}$ and ${\hat{H}}_{}^{}$ have the
same spectrum $E^{}_n$. This fact indicates that the two operators
are equivalent,
\begin{equation}\label{11}
{\hat{H}}_{}^{}={\hat{U}} {\hat{H}}_{}^{(0)}{\hat{U}}_{}^{\dag},
\end{equation}
where ${\hat{U}}$ is a unitary operator. Indeed, a unitary
transformation preserves the spectrum in a new orthonormal basis
${\hat{U}}{\psi}^{}_{n{\nu}}$.

Unitary matrix connecting two-band operators (\ref{8}) and
(\ref{9}) depend on three angle variables, $\vartheta$, $\theta$,
$\varphi$. Matrix elements can be presented in the form
$U^{}_{11}=U^{\ast}_{22}=e_{}^{i{\vartheta}}{\cos}{\varphi}$,
$U^{}_{12}=-U^{\ast}_{21}=e_{}^{i{\theta}}{\sin}{\varphi}$. The
explicit form of these matrix elements can be directly obtained
from Eq. (\ref{11}).

One can see that the excitonic insulator with toroidal magnetic
order described by the approximate Bloch Hamiltonian (\ref{8}) and
the Haldane model \cite{Haldane} described by the Bloch
Hamiltonian (\ref{9}) are equivalent that is can be related to a
class of models in which the Haldane state can be apparent.

To create the Haldane state one needs only the time-reversal
symmetry break in 2D crystal. Therefore, it seems that artificial
heterostructures in the form of quantum well systems may be the
materials in which such a state can be discovered. To obtain
desired properties of the electron spectrum of the heterostructure
one can easily correct this spectrum by external in-plane magnetic
field \cite{GKK} or by varying of magnetic impurity concentration.
We believe that precisely 2D quantum wells with toroidal magnetic
order are the most suitable systems to realize the Haldane state.

\vspace{0.6cm}

We would like to thank G.E.~Volovik for drawing our attention to
an analog of the QHE state without external magnetic field that
should be apparent in $^3{\text{He}}-A$ and multi-band insulators
with spontaneous magnetic moment studied in his early papers Refs.
\cite{V1,V2}. The work was supported, in part, by the Russian
foundation for basic research, project N 09-02-00682-a.

\end{document}